# Strong substrate influence on atomic structure and properties of epitaxial VO$_2$ thin films

*Atul Atul\*[a], Majid Ahmadi [a], Panagiotis Koutsogiannis [a,b], Heng Zhang [a], Bart J. Kooi\*[a]*

[a]Zernike Institute for Advanced Materials, University of Groningen, Nijenborgh 4, 9747 AG, Groningen, The Netherlands
[b]Zumalagarregui 23 2B, Zaragoza 50006, Spain

**ABSTRACT**

The metal-insulator transition (MIT) observed in vanadium dioxide (VO$_2$) has been a topic of great research interest for past decades, with the underlying physics yet not fully understood due to the complex electron interactions and structures involved. The ability to understand and tune the MIT behaviour is of vital importance from the perspective of both underlying fundamental science as well as potential applications. In this work, we use scanning transmission electron microscopy (STEM) to investigate cross-section lamella of the VO$_2$ films deposited using pulsed laser deposition (PLD) on three substrates: c-cut sapphire, TiO$_2$(101) and TiO$_2$(001). Advanced STEM imaging is performed in which also the oxygen atom columns are resolved. We link the overall film quality and structures on atomic and nanoscale to the electrical transition characteristics. We observe poor MIT characteristics on c-sapphire due to the presence of very small domains with six orientation variants, and on TiO$_2$ (001) due to the presence of cracks induced by stress relaxation. However, the MIT on TiO$_2$ (101) behaves favourably, despite similar stress relaxation which, however, only lead to domain boundaries but no cracks.

# INTRODUCTION

Metal-insulator transitions (MITs) have been extensively studied for decades owing to their importance in fundamentals of condensed matter physics[1,2] as well as for their (opto)electronic applications.[3–6] The most famous transition metal oxide showing MIT is vanadium dioxide ($VO_2$), which exhibits a first-order transition in bulk form just above room temperature, at 340 K,[7,8] from a low temperature insulating phase to a high temperature conducting phase. This is accompanied with a structural phase transition (SPT) from monoclinic ($M_1$) crystal with space group 14 (P 21/c) to a tetragonal (rutile-R) crystal with space group 136 (P 42/mnm).[9,10] Due to the complex nature of the correlated electron interactions including a variety of intermediate states,[11–14] the transition is still not fully understood.[15–18] $VO_2$ is considered as a promising candidate for oxide-based electronic applications[19,20] due to its excellent transition characteristics of ultrafast resistivity change of over four orders of magnitude with a sharp hysteresis.[1,21] This MIT behaviour can be further tuned by external stimuli like electric and magnetic field, light, doping, and strain.[22–27]

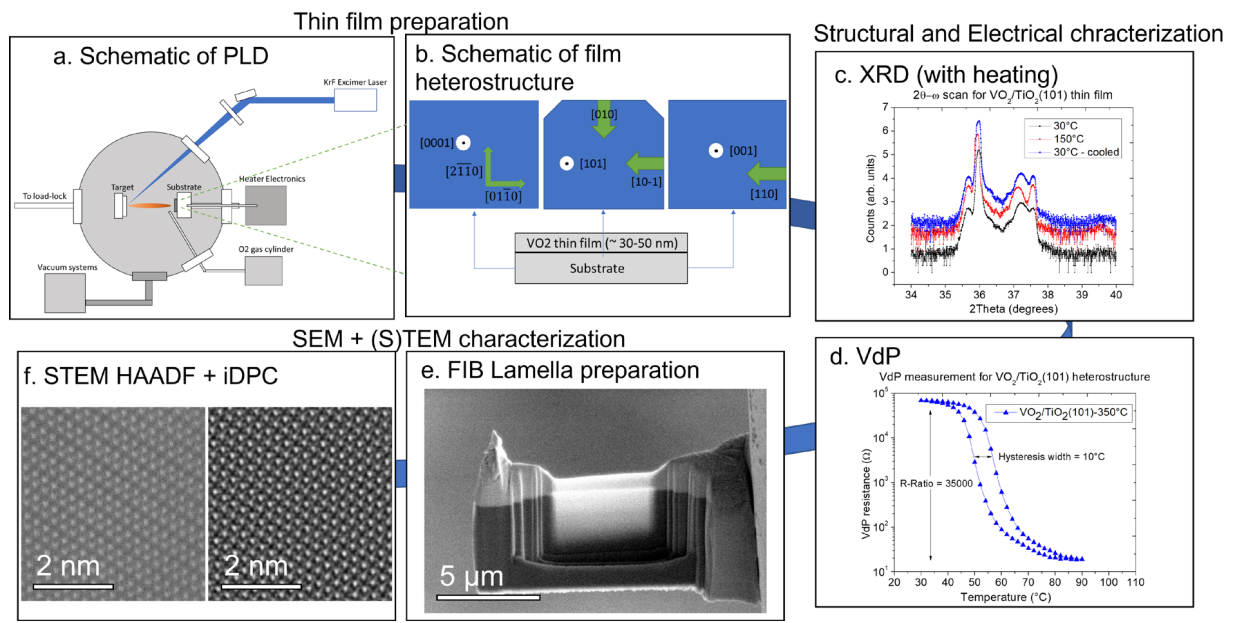

**Figure 1**. Schematic representation of the experiments involved in the present work. Using (a) pulsed laser deposition (PLD), (b) epitaxial $VO_2$ films are grown on c-sapphire, $TiO_2$(101) and $TiO_2$(001) substrates. They are characterized using (c) X-ray diffraction for structure, and (d) Van der Pauw for sheet resistivity measurements across the transition. (e) Cross-section lamella from the three films are made using a focused ion beam system, and these lamella are analyzed using (f) advanced scanning transmission electron microscopy (STEM).

$VO_2$ thin films have been previously grown using diverse techniques, both physical and chemical, on a variety of substrates.[28–32] For such epitaxially grown films, the nature of the microstructure, including distribution of differently oriented grains and their grain boundaries, strongly affect the transition properties.[33–35] Such studies have been performed individually on the most common substrates used for growing $VO_2$ – sapphire[36–42] and $TiO_2$,[43–47] but we could not find any studies of such films and their microstructure across substrates using very similar growth conditions to maximize their comparison. Furthermore, the microstructure has been commonly studied using diffraction and in fewer cases with direct imaging using scanning transmission electron microscopy (STEM). However, then only the high-angle annular dark field (HAADF) mode was used that does not allow imaging of the O atomic columns next to the heavier V atomic columns. Another challenge is that the microstructure is not fully resolved using only plan-view studies, and cross-section preparation (e.g. by focused ion beam (FIB)) often lead to surface amorphization and damage making proper analysis difficult, especially when combined with the presence of features like domain boundaries and cracks in the films.[46]

Here we study the transition properties of $VO_2$ films grown on different substrates using pulsed laser deposition (PLD) under similar growth conditions. We correlate the differences in the MIT behaviour, as seen from Van der Pauw electrical measurements, to the structures obtained from X-ray diffraction (XRD), which show the significantly different strain profiles of the films. To understand the origin of these differences at the nanoscale, we make cross-section lamella along non-equivalent orientations from the films which are thinned at temperatures well below 0 °C to prevent damage and amorphization, and then study them using STEM, combining the common HAADF imaging mode with integrated differential phase contrast (iDPC)-STEM imaging[48] where the oxygen atomic columns are also well resolved. The basic scheme of experiments involved in the present work is illustrated in Figure 1. The combination of the experiments described here give an unprecedented detailed and comprehensive picture of the morphology of the films on various substrates and its relation to the MI transition properties, which is of vital importance when making devices based on these films.

# RESULTS AND DISCUSSIONS

## Structural and Electrical Characterization

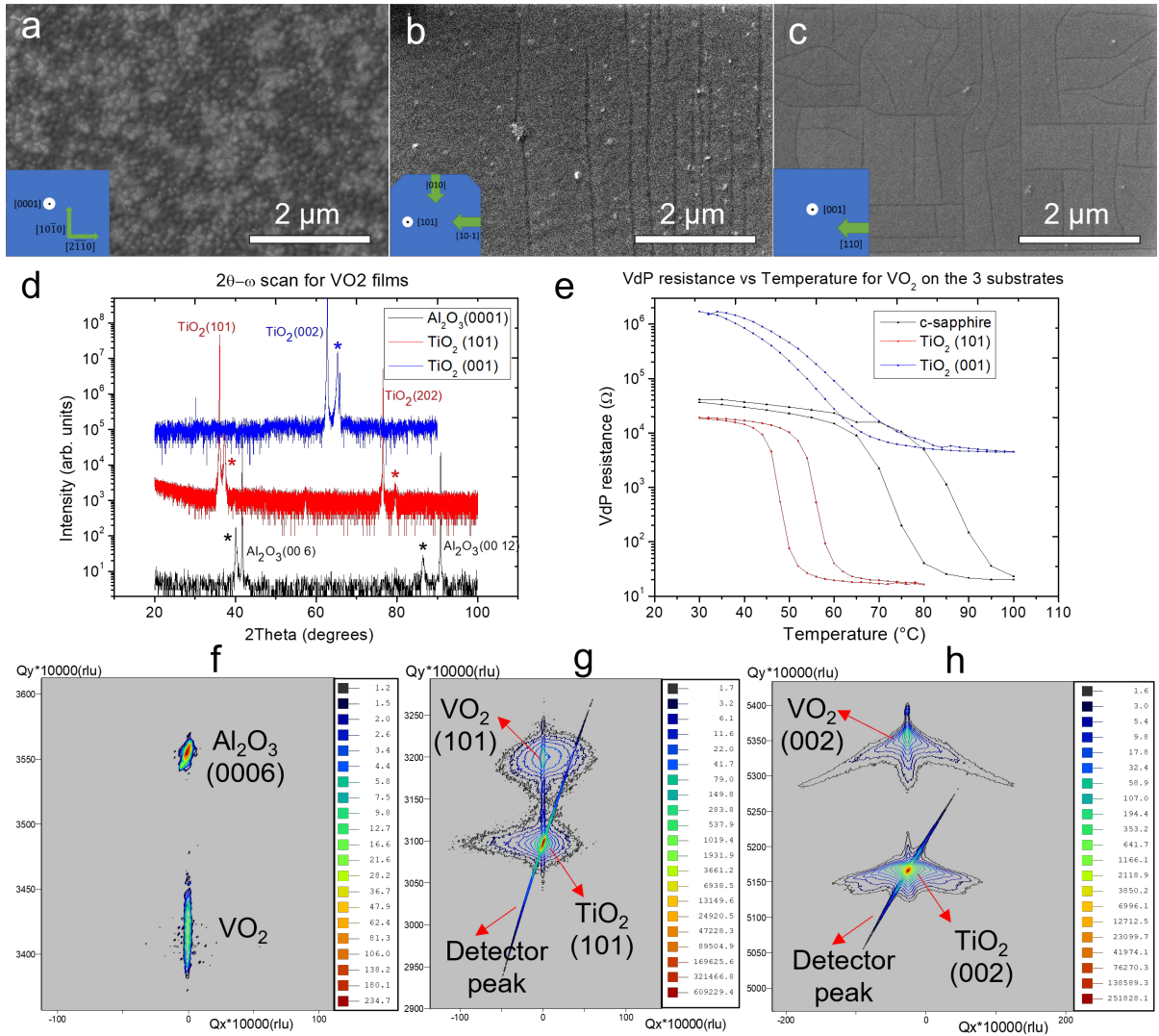

**Figure 2**. SEM image of the VO$_2$ surface on (a) c-cut sapphire (b) TiO$_2$(101), and (c) TiO$_2$(001) with orientation of the substrates indicated by the insets, (d) XRD 2θ-ω scans of the films with vertical offsets (with peaks from VO$_2$ films marked with *), and (e) Sheet resistance of the films as a function of temperature measured using Van der Pauw technique, Reciprocal space maps of (f) VO$_2$ on c-sapphire around (0006) peak of Al$_2$O$_3$, (g) VO$_2$ on TiO$_2$(101) around (101) peak of TiO$_2$, and (h) VO$_2$ on TiO$_2$(001) around (002) peak of TiO$_2$.

In Figure 2 (a-c), we show scanning electron microscopy (SEM) images of the surfaces of VO$_2$ thin films deposited on c-cut sapphire, TiO$_2$(101) and TiO$_2$(001) substrates (taken with Ion Conversion and Electron (ICE) detector), with the schematic orientations of the substrates given in the insets. From the SEM images, we can see that VO$_2$ on sapphire has very small grains (<100 nm), whereas the VO$_2$ films on both TiO$_2$ substrates do have domains on µm length scale. On TiO$_2$(101), the domain boundaries are all perpendicular to the (010) edge plane. This leads to high aspect ratio domains with widths <1μm, but being very long parallel to the boundaries. On $TiO_2$(001), the domain boundaries are perpendicular to both {110} edge planes which give individual grains of variable lengths and widths, but one of these dimensions is usually of the order of ~0.5-1 μm.

The X-ray diffraction (XRD) 2θ-ω scans of the films with 2θ values in the range 20° to 100° and intensity in log scale are shown in Figure 2d with vertical offsets between the scans for clarity. $VO_2$ on sapphire has a broad peak near 40.1° close to the very sharp (0006) peak of the substrate at 41.68°, and similarly a broad peak at 86.4° close to (00 12) peak of c-$Al_2O_3$ at 90.818°. The peak has previously been reported as being either (020) or (002) of monoclinic (M1) $VO_2$[49–51], as both of them give very similar 2θ values, which implies that the film as-cooled down after high temperature growth has either b- or c- axis out-of-plane, respectively. It is not possible to differentiate between those peaks with 2θ-ω scans or even reciprocal space maps.[51] On $TiO_2$(101), there is a sharp M1 $VO_2$(101) peak at 37.31° close to the (101) substrate peak at 36.08°, and M1 $VO_2$(202) peak close to (202) substrate peak at 76.53°.[52] The reciprocal space map around the (101) peak in Figure 2g shows us the in-plane epitaxial relationship with strain around both the film and substrate peaks. On $TiO_2$(001), a sharp substrate peak (002) at 62.8° can be seen, with a major film peak at 65.3° with a shoulder peak at 65.8°. The reciprocal space mapping around the (002) peak also shows the in-plane orientation matching, but with a significant strain (gradient) component of the film peak which is an indication of an interface layer.[53,54] Overall, the XRD results clearly demonstrate the highly epitaxial nature of the films.

The results of Van der Pauw (VdP) electrical measurements in Figure 2e highlight the effects of these structural differences and in particular, the effect of the domain boundaries on the electrical properties. The VdP results for all the samples on the three substrates are provided in Supplementary Figure S2, with all the transition parameters in Supplementary Table 2. The samples chosen for comparison in Figure 2e are all made using 20000 pulses under oxygen pressure of 7.5 mtorr, with the sapphire substrate kept at 500 °C and the $TiO_2$ substrates at 375 °C, but similar trends are seen with other samples. The films on all the substrates exhibit an insulator-to-metal (or semiconductor) transition upon heating and return to the original state upon cooling. However, whereas the sheet resistance of the high-temperature state is of the order of $10^1$ Ohms for films on c-sapphire and $TiO_2$(101), it is ~$10^3$ Ohms or higher for films on $TiO_2$(001) which suggests that the low-resistance state on $TiO_2$(001) is still a semi-conducting state instead of being the expected metallic one. Furthermore, there are also notable differences in the shape of the hysteresis curves of the samples on the different substrates. The samples on $TiO_2$(101) exhibit high quality transition characteristics, with a high resistance ratio of ~$10^3$ with a hysteresis width of less than 10 °C. The transitions on sapphire and $TiO_2$(001) show similar resistance ratios, but occur over a much wider temperature range, greater than 30 °C, even if the hysteresis width, measured as the difference in temperatures where the derivatives of resistance versus temperature show extrema, is of the same order for all three substrates. A plausible explanation could be the differences in the domain structures. For instance, the domains on TiO$_2$(001) could individually have a sharp transition, but the presence of multiple differently oriented domains smear out the transition over a wide temperature range. Domains, all being oriented along the same direction on TiO$_2$(101) substrate then leads to a significantly sharper transition. However, this reasoning clearly does not extend to c-sapphire which has a finer domain structure than TiO$_2$ visible in SEM images, but still has a wider transition than TiO$_2$(101). To further elucidate the reasons for these differences (in Fig. 2e), we performed a comprehensive STEM analysis from cross-section lamellas prepared from these films, which will be discussed below.

## VO$_2$ on sapphire

Cross-section lamellas were prepared parallel to the short and long sides of the substrate with the directions schematically shown in Figure 2a. In Figures 3a and 4a, we show the selected area electron diffraction (SAED) patterns obtained from the two lamella with the reciprocal lattice vectors of the substrate and film in blue and red color, respectively. The diffraction spots were indexed using the c-sut sapphire as calibration reference, specifically using the (0006) peak of Al$_2$O$_3$. In Figure 3a the (0006) spot of Al$_2$O$_3$ is parallel to the VO$_2$ spot at 0.441 Å$^{-1}$, which is the (002) peak of monclinic (M1) VO$_2$ as discussed before. This gives us $(000\bar{6})_{Al2O3}\|(002)_{VO2}$, and the presence of the $(\bar{2}110)$ and $(\bar{2}11\bar{6})$ diffraction spots gives us the zone-axis $[01\bar{1}0]$. Overall, we get $[01\bar{1}0]_{Al2O3}\|[100]_{VO2}$ and $(000\bar{6})_{Al2O3} \| (002)_{VO2}$ as the orientation relationship between the film and substrate. However, note that the d-spacings of (020) and (002) are very similar, so that the orientation could also be written as $[01\bar{1}0]_{Al2O3}\|[100]_{VO2}$ , and $(000\bar{6})_{Al2O3} \| (020)_{VO2}$. Similarly, the $(000\bar{6})$ and $(01\bar{1}2)$ spots in Figure 4a give us, as expected, the orthogonal zone-axis $[2\bar{1}\bar{1}0]$ for the 2$^{nd}$ lamella,which is in agreement with previous work.[51] Overall, we get $[2\bar{1}\bar{1}0]_{Al2O3} \| [010]_{VO2}$ , and $(000\bar{6})_{Al2O3} \| (002)_{VO2}$ as the orientation relationship between the film and substrate. Since the c-axis of sapphire (space group R-3c) is a three-fold axis, this means that the film is composed of three distinct domains with mutual in-plane rotation[55,56] around the two growth axes (which are all shown in Supplementary Figure S6) giving a pseudo six-fold symmetry in the diffraction pattern. Further investigations were done based on atomic resolution STEM analysis using both High-Angle Annular Dark Field (HAADF) and Integrated Differential Phase Contrast (iDPC) images and also closely comparing experimental and simulated images.

Overview STEM HAADF and iDPC images of the two lamellas with orthogonal orientations are shown in Figure 3b-c and Figure 4b-c, respectively. From these images, domains with different orientations can be readily identified, and from individual domains images are cropped and shown in Figures 3d-o and Figures 4d-g.

The domains have a columnar shape with a diameter not exceeding 10 nm and have a nearly uniform width over the entire film thickness. From lamella prepared along the $[01\bar{1}0]_{Al2O3}$ projection axis, three distinct domains in M1 $VO_2$ have been identified.

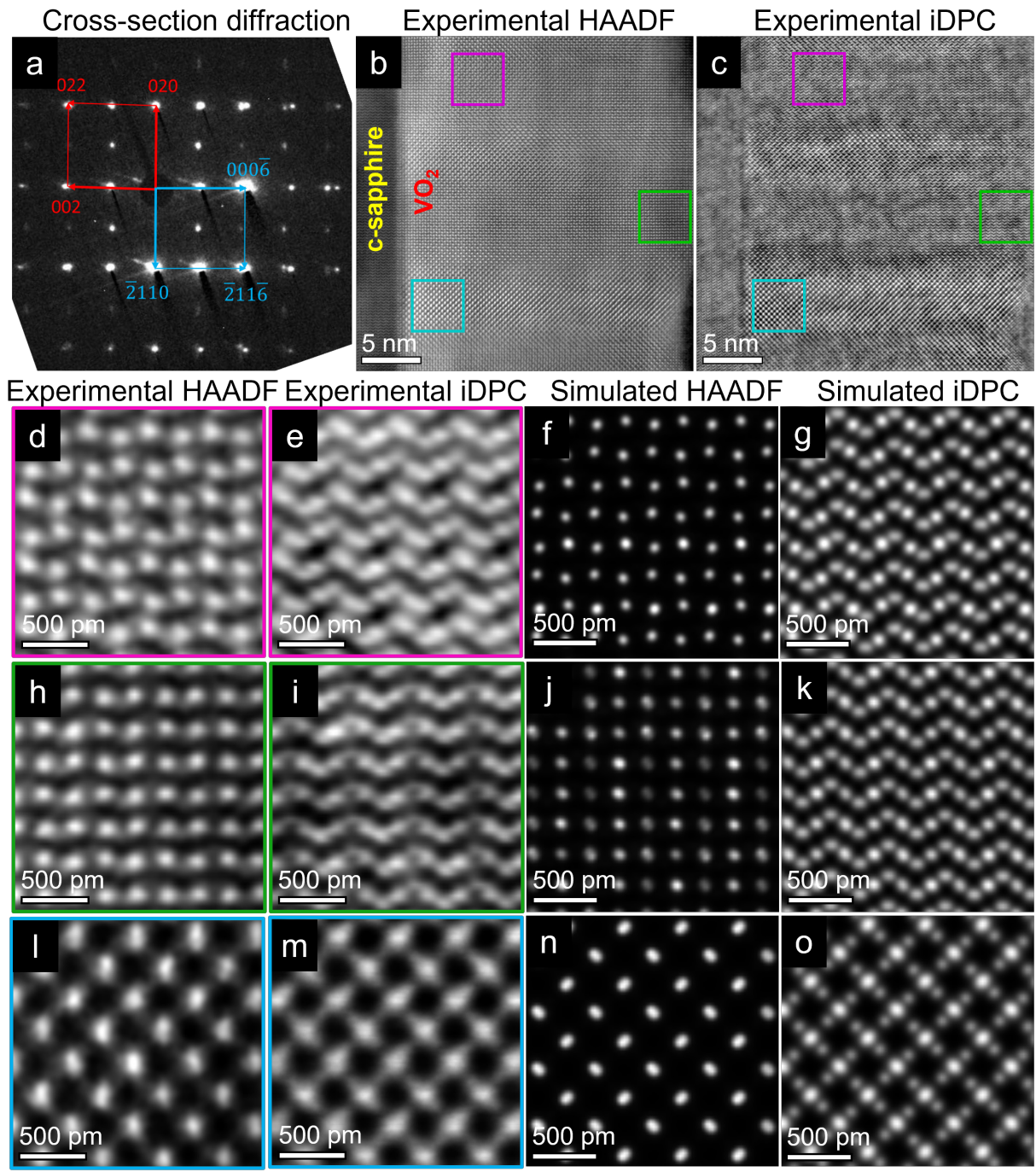

**Figure 3**. (S)TEM HAADF and iDPC images taken at 300 kV, and simulations for M1 $VO_2$ on c-cut sapphire. (a) Diffraction pattern of the lamella prepared parallel to the long side of the substrate, (b,c) HAADF, iDPC image for lamella with $[01\bar{1}0]$ projection axis of c-cut sapphire, with $VO_2$ domains of different orientations identified by comparing to corresponding simulations. Individual grains are cropped for clarity. The HAADF and iDPC of different grains with their simulations are (d-g) along [101], (h-k) along [120], and (l-o) along [100] of $VO_2$.

The simplest domain of these three is shown in Figure 3l-m, with V-V distances of 0.336 nm. This is easily identified as the [100] zone axis from the symmetry of the V-V distances which are 0.328 nm in the corresponding simulations in Figure 3n-o. The difference with the distances seen experimentally could be due to the domain being very close to the interface where it has high strain due to the lattice mismatch. This domain orientation is possible with either b or c out-of-plane growth axis, as seen from Figure S5.

Figures 3d-e show images of a domain with alternating short and long bonds parallel to the interface with V-V bond distances of 0.298 nm and 0.218 nm for long and short bonds respectively. Perpendicular to the interface there is a zig-zag pattern observable in the iDPC image, i.e. when including the oxygen atomic columns. This matches well to the model of [101] zone axis, and the corresponding simulations in Figure 3f-g clearly show the long and short bonds being 0.296 nm and 0.223 nm parallel to the interface and the zigzag pattern in the iDPC image perpendicular to the interface. So, this domain has grown with the b-axis out-of-plane.

Another zone axis, shown in Figure 3h-i, does not produce a structure having long and short alternating V-V bonds in either of the directions parallel and perpendicular to the interface (see Fig. 3h). Still, there is a clear zig-zag pattern perpendicular to the interface present in the iDPC image (Fig. 3i). The observed structure is very close to the model of the [120] zone-axis. Simulated HAADF and iDPC images for this zone-axis, shown in Figures 3j and 3k, respectively, agree well with the experimental images. So this domain has the degenerate [120] or [1$\bar{2}$0] orientation, both of which have c-axis out-of-plane growth. The above results thus show that we have multiple domains with either b- or c-axis out-of-plane growth within the same film.

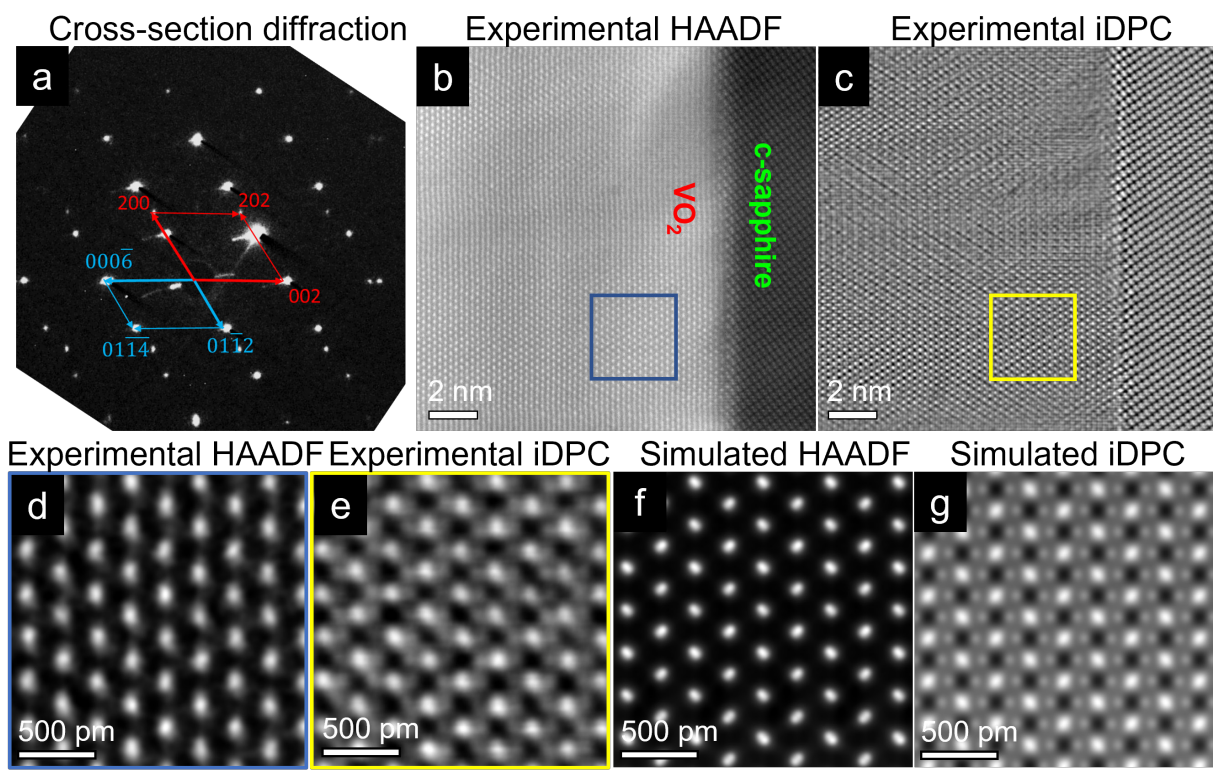

**Figure 4.** (S)TEM HAADF and iDPC images taken at 300 kV, and simulations for monoclinic VO$_2$ on c-cut sapphire. (a) Diffraction pattern of the lamella prepared parallel to the short side of the substrate, (b,c) HAADF, iDPC image for lamella with $[2\bar{1}\bar{1}0]$ projection axis of c-cut sapphire. HAADF and iDPC of M1 VO$_2$ [102] domain has been cropped in (d-e), identified by comparing to the simulations shown in (f-g).

From the lamella prepared with $[2\bar{1}\bar{1}0]_{Al_2O_3}$ projection axis, only one monoclinic VO$_2$ domain orientation could be clearly identified as shown in Figure 4d-e, which has a hexagonal arrangement of O atoms around V. There are two possible zone-axes with such an arrangement of atoms, namely [010] with c-growth axis and [102] with b-growth axis. However, with the [010] zone-axis, the V atomic columns would form dimerized planes, i.e. with alternating short and long interplanar distances, which is not seen in this image. So the orientation must be [102] for which the simulations are shown to be in close agreement in Figure 4f-g. Apart from these domains which could be confirmed by comparing the V-V or V-O distances in the simulations, there were other domains, whose images are shown in Supplementary S7, which are significantly harder to verify since the V and O atomic columns are very close to each other such that they do not provide distinctly resolved signals in iDPC images.

An important observation from the images is that the domains are very small (often smaller than 5-10 nm) with no sharp boundaries, which means that many domains overlap in the image along the projection direction of the STEM imaging. The lamella have thicknesses of say 60 nm and then we would readily have six or more overlapping domains in TEM images, which would make it impossible to image any domain individually. In STEM, since the focusing is more confined at a certain depth

(due to the 'waist' of the focused electron beam) particularly since we can employ a large convergence angle in aberration-corrected STEM, it is still possible to distinguish some domains individually like we demonstrated above. Still, even in this case the very small domain size often leads to overlapping domains, making further analysis of these films very complex. Nevertheless, the presence of such small domains combined with the two possible out-of-plane axes clearly explains why the MI transition is spread over a large temperature interval as shown in Figure 2e.[57]

**VO$_2$ on TiO$_2$(101)**

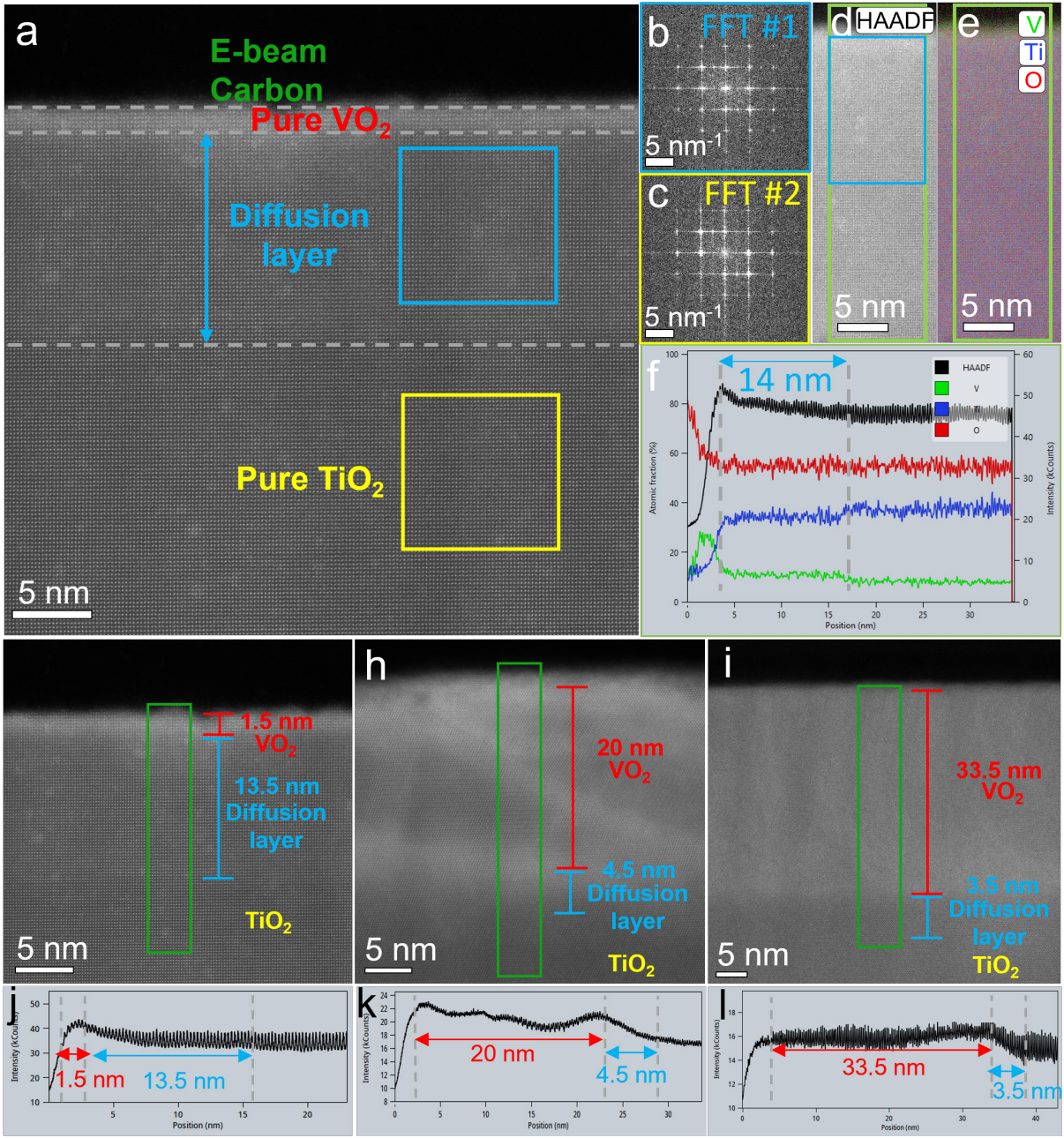

**Figure 5**. STEM HAADF images of $VO_2$ on $TiO_2$(101) grown at varying substrate temperatures from 500 °C to 350 °C, taken at 300 kV. (a) $VO_2$ on $TiO_2$(101) grown at 500 °C viewed along $[\bar{1}01]$ direction, (b-c) FFT of the intermixing region and pure $TiO_2$ region. The regions are confirmed using an EDS scan with the (e) elemental mapping of a (d) selected area, with (f) its profile showing the variation of the different elements along the line scan. (g-i) HAADF images of samples grown at 500 °C, 400 °C and 350 °C, and (j-l) respective profiles of the line scans.

Cross-section lamella for samples on $TiO_2$(101) were made with the long side of the lamella parallel to the (010) edge plane for STEM analysis of the sample which means perpendicular to the domain boundaries (such that the boundaries can be observed edge-on in the cross-section). In Figure 5a, we show the STEM HAADF image for the case where the $VO_2$ film was grown on $TiO_2$ at 500 °C. The contrast difference observed in the HAADF image shows three regions. A very thin (~2nm) top layer is very distinct, because it is clearly brighter than the regions below. Still, there is a ~14nm middle layer that is marginally brighter than the $TiO_2$ substrate at the bottom, which is observed along the [010] zone-axis as confirmed by the FFT in Figure 5c. In order to understand the origin of the three regions better, we performed EDS mapping of a selected area to capture these regions as shown in Figure 5d-e, with the elemental profiles of the line scan in Figure 5f. Note that the absolute values for atomic fractions in Figure 5f are generally not accurate, mainly due to spurious X-rays, but that the linescans still provide reliable qualitative information. In particular the three different regions can now be observed more distinctly, i.e. a pure $VO_2$ top layer of ~2 nm thick, followed by a ~14 nm thick middle layer where V has been diffused into the $TiO_2$ and finally the (pure) $TiO_2$ substrate at the bottom. The O percentage stays constant over the entire line scan which indicates that the three layers all have composition $MO_2$, where M is either V, Ti or some intermixed $V_xTi_{1-x}$. This intermixing layer also has the same (tetragonal) structure as $TiO_2$ as confirmed from the FFT in Figure 5b. Important to note that the intermediate layer does not show a typical diffusion profile where the V concentration is highest at the top and gradually reduces to zero inside the $TiO_2$ substrate. In contrast, the intermediate layer has relatively constant V to Ti concentration ratio (roughly being 0.1) and this $V_xTi_{1-x}O_2$ also shows a rather distinct interface with the pure $TiO_2$ substrate.

This sample with the about 2 nm thick $VO_2$ top layer did not show any MIT due to the too limited layer thickness of $VO_2$. In order to reduce the thickness of the intermediate layer and to get thicker pure $VO_2$ films, the growth temperature was reduced in steps from 500 °C down to 350 °C. The HAADF images in Figure 5g to 5i confirm that this growth strategy works. Below a growth temperature of 400 °C, the intermediate layer is ~3-4 nm thick and further reduction in growth temperature leads to an increase in $VO_2$ film thickness, but with no significant reduction in intermediate layer thickness. The films grown in the range 350-450 °C exhibited excellent MIT characteristics as shown in Supplementary S2 and discussed earlier. Overall, we see considerable V-Ti intermixing over the thickness range from ~3 nm to ~15 nm, which provides us with a new approach to make these superlattice structures by growing the VO$_2$ films at controlled temperatures in the range 350-450 °C. Further studies are required to study the properties of these heterostructures since the MIT characteristics should be affected by this intermediate intermixing layer.[45]

From a sample grown at 375 °C we produced multiple lamellas from areas with different domain widths. In Figure 6a, we show an overview drift-corrected (DCFI) image of a lamella taken with a camera length of 190 mm for better contrast. This allows us to clearly distinguish the (edge-on observed) domain boundaries, which are typically separated by a few 100nm to ~1μm. A HAADF image of a domain boundary region is shown in Figure 6b. Focusing on the TiO$_2$ substrate, we see that its orientation is as expected [010] from the FFT in 6c. We see through FFTs in Figure 6d-f that the domain boundary region in fact contains two closely spaced parallel domain boundaries, i.e. that what looks at a distance like a single domain boundary is in fact a thin lamella of a domain, where its two boundaries make an angle of about 50 degrees with the interface and surface. Note that this orientation is always inclined to one side and that the lamellae do not run at plus and minus 50 degrees through the film. This asymmetry agrees with the asymmetry induced by the substrate TiO$_2$(101) where in this case the c-axis is oriented with a component to the left and the a-axis with a component to the right (see the indicated c- and a-axis in Fig. 6b). Now the FFT analysis shows that the thin VO$_2$ lamella is viewed along the $[010]_{M1}$ zone axis (FFT #3 in Fig. 6) and that on both sides of the thin lamella the orientation of the two domains are actually identically corresponding to the $[102]_{M1}$ zone axis (FFT #2 and #4 in Fig. 6). These structural differences are also observable more clearly in the DCFI-HAADF image shown in Supplementary S8(a). Note that even a more intricate structure is possible where the domain in the thin lamella is twinned with a boundary that runs perpendicular to the interface and surface of the film. So, in this case next to the $[010]_{M1}$ domain, also the $[0\bar{1}0]_{M1}$ domain is present, which has opposite orientation of the dimerized V-V planes in the monoclinic VO$_2$ structure. One such domain boundary region with both $[010]_{M1}$ and $[0\bar{1}0]_{M1}$ grains inside the thin lamella is also shown in Supplementary S8(b). These degenerate structures are equivalent to the non-degenerate original $[100]_R$ and $[\bar{1}00]_R$ orientations, which would obviously still be fully epitaxial with the tetragonal substrate. Obviously these domain boundaries are formed when the sample is being cooled down after growth, because the VO$_2$ grows fully epitaxially on the TiO$_2$ at the growth temperature where the VO$_2$ still has the same tetragonal structure as the TiO$_2$ substrate. However, when the VO$_2$ transforms during cooling from tetragonal to monoclinic the symmetry breaking causes the domain formation. Another important observation is that, although from the top-view of the sample the domain boundaries run perpendicular to the TiO$_2$(010) edge plane and seem parallel to the TiO$_2$(101) edge plane, the boundary plane is running inclined through the thickness of the film. This also explains the extended width of the domain boundaries when observed in the plan-view SEM image in Figure 2b.

In Figure 7a-b, we also show high-resolution HAADF and iDPC images of the sample taken at some distance away from domain boundaries, from which regions of VO$_2$ and TiO$_2$ have been cropped in Figure 7c-d and e-f, respectively. From the HAADF image, we can see that the film has excellent orientation matching with the substrate, and that the domains are essentially single-crystals. From the iDPC images, we see a hexagonal-like arrangement of O atoms around the central V or Ti atoms with the two V-O bonds perpendicular to c-axis being equal, but differ from the other four V-O bonds which are equal to each other. The observed atomic structures are in excellent agreement with the simulations done for the monoclinic (M1) [102]$_{VO2}$ and the tetragonal [010]$_{TiO2}$ orientations shown in Figure 7g-j. In summary, when we view for the TiO$_2$(101) substrate along TiO$_2$[010], [102] domains are dominantly present and in between them a thin inclined domain of [010] and/or [0$\bar{1}$0] type is present.

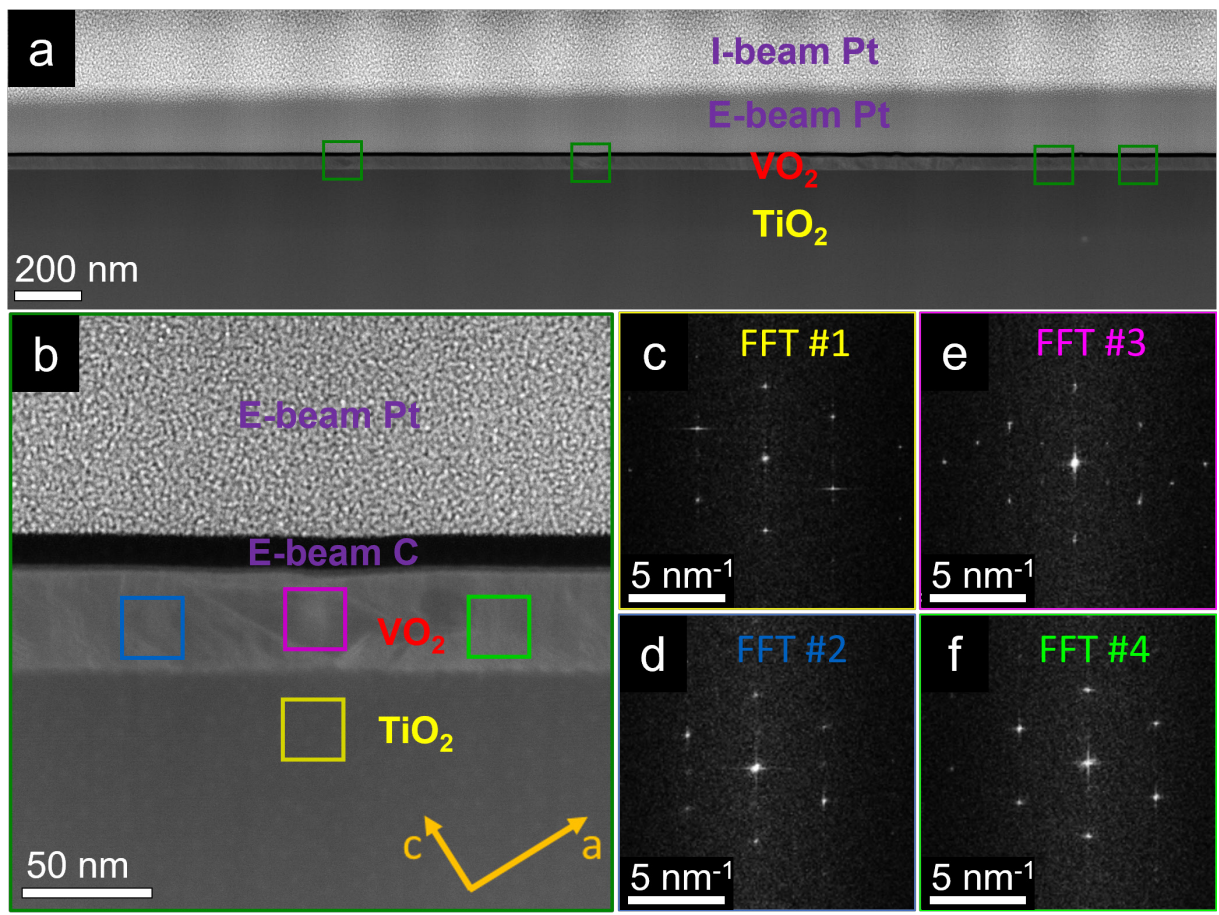

**Figure 6**. STEM-HAADF images of VO$_2$ on TiO$_2$ (101) substrate grown at 375 °C, taken at 300 kV. (a) Overview HAADF-DCFI image showing the domain boundaries, one of which is shown in (b). The structures inside and outside the domain boundary are marked in, with FFTs 2-4 shown in (d-f), compared to the TiO$_2$ substrate with its FFT shown in (c).

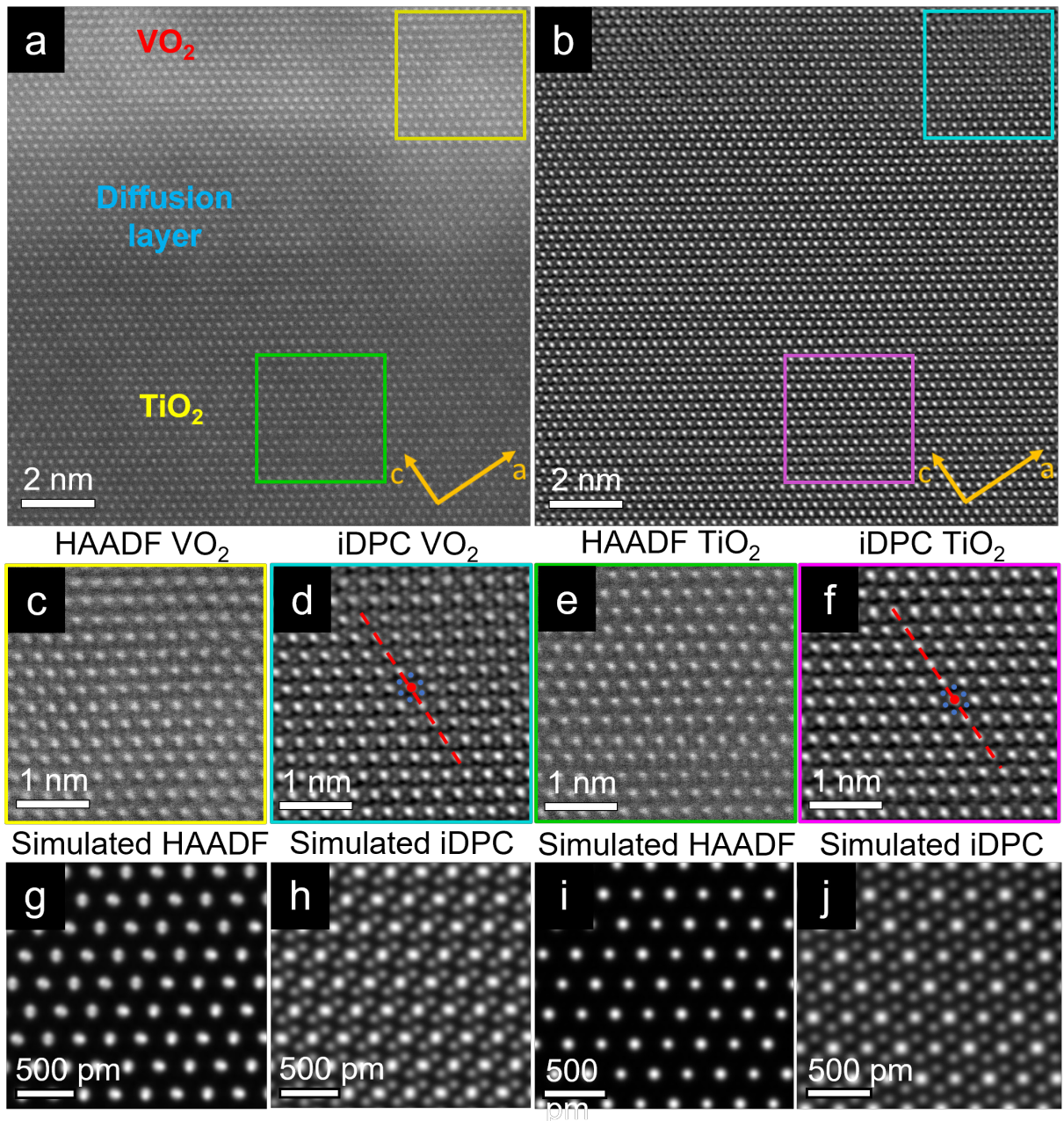

**Figure 7**. STEM HAADF and iDPC images of VO$_2$ on TiO$_2$(101) substrate grown at 375 °C, taken at 300 kV. (a-b) HAADF and iDPC of the lamella away from domain boundaries, with cropped regions (c-d) VO$_2$ (e-f) TiO$_2$ substrate for clarity showing epitaxial film (Red lines are to mark the c-axis in the image). Simulations of HAADF and iDPC for (g-h) VO$_2$ with [102] orientation, (i-j) TiO2 with [010] orientation.

## VO$_2$ on TiO$_2$(001)

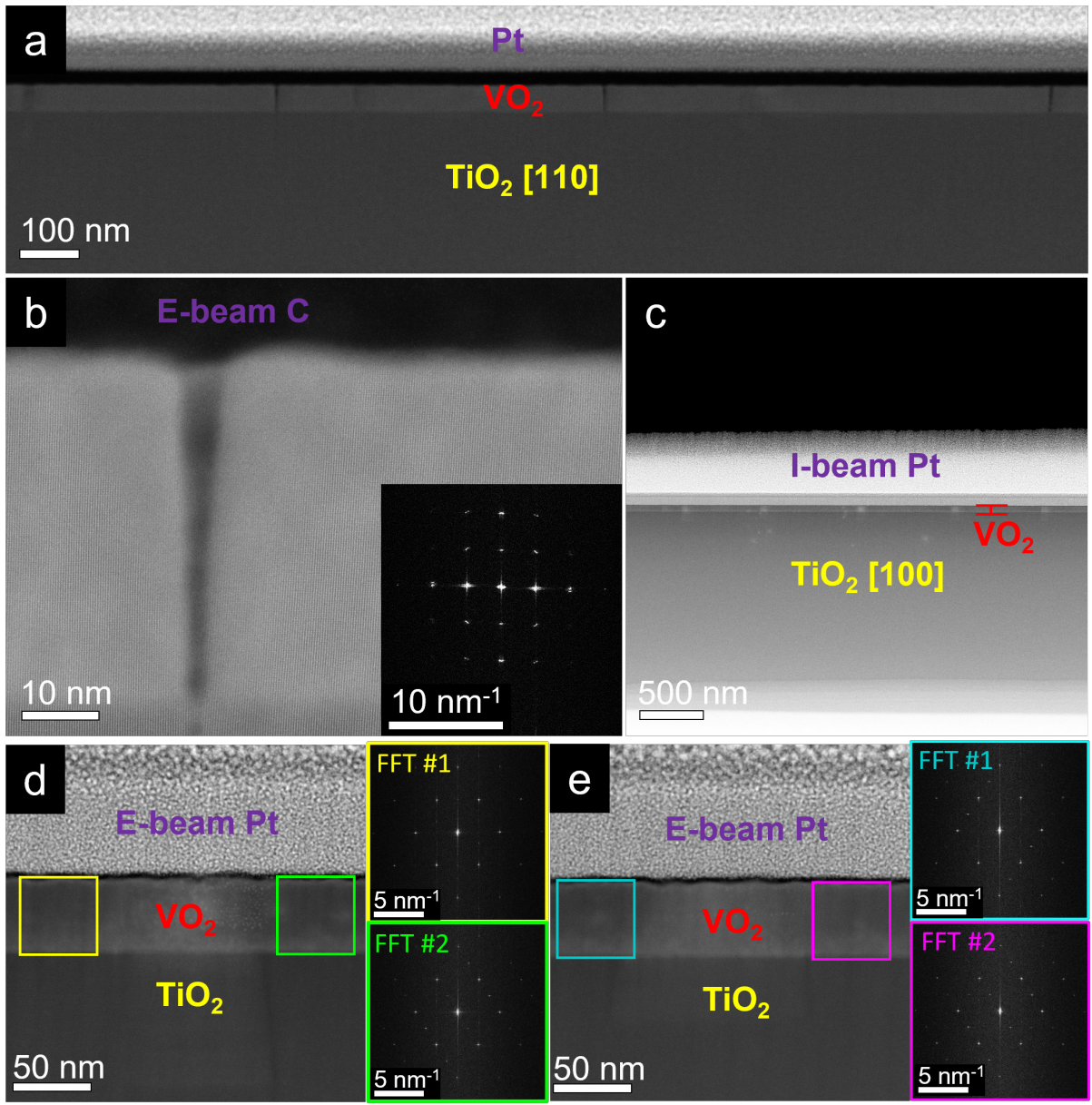

**Figure 8**. (a) Overview (DCFI) HAADF image of cross-section of VO$_2$ on TiO$_2$ (001) grown at 375 °C along [110] projection which shows the features are actually cracks, (b) HAADF of a crack with inset FFT, (c) Overview of cross-section taken along [100]/[010] projection, (d-e) HAADF of two of these domain boundaries where domains with different orientations (shown in the inset FFTs of the regions) meet.

From Figure 2, we know that on TiO$_2$(001), the domain boundaries are parallel to the two orthogonal {110} edge planes of the substrate. Cross-section lamella were made from VO$_2$ grown at 375 °C, both parallel to one of the two edge planes and at the intermediate angle of 45° (rotation around out-of-plane TiO$_2$ c axis). In Figure 8a, we show an overview image of the lamella made parallel to the edge plane, so the zone-axis is [110]. We see multiple dark line running across the VO$_2$ film perpendicular to the VO$_2$/TiO$_2$ interface. These dark line are about 0.5-1 μm apart from each other which coincides well with the domain sizes observed in the SEM image in Fig. 2c. However, when zooming in on the dark line in Figure 8b, it becomes obvious that the lines are not domain boundaries, but rather cracks developed in the film, which even can propagate considerably into the substrate (see Supplementary S9). The FFT inset in the figure shows that the two sides of the boundary still have $VO_2(1\bar{1}0)$ planes parallel to the crack boundary, but they have a small mutual rotation (around [110]) from each other due to the crack opening. These cracks have been reported to be thermally induced - when the sample cools down through the transition temperature, the cracks develop as a means of strain relaxation.[43,44,58]

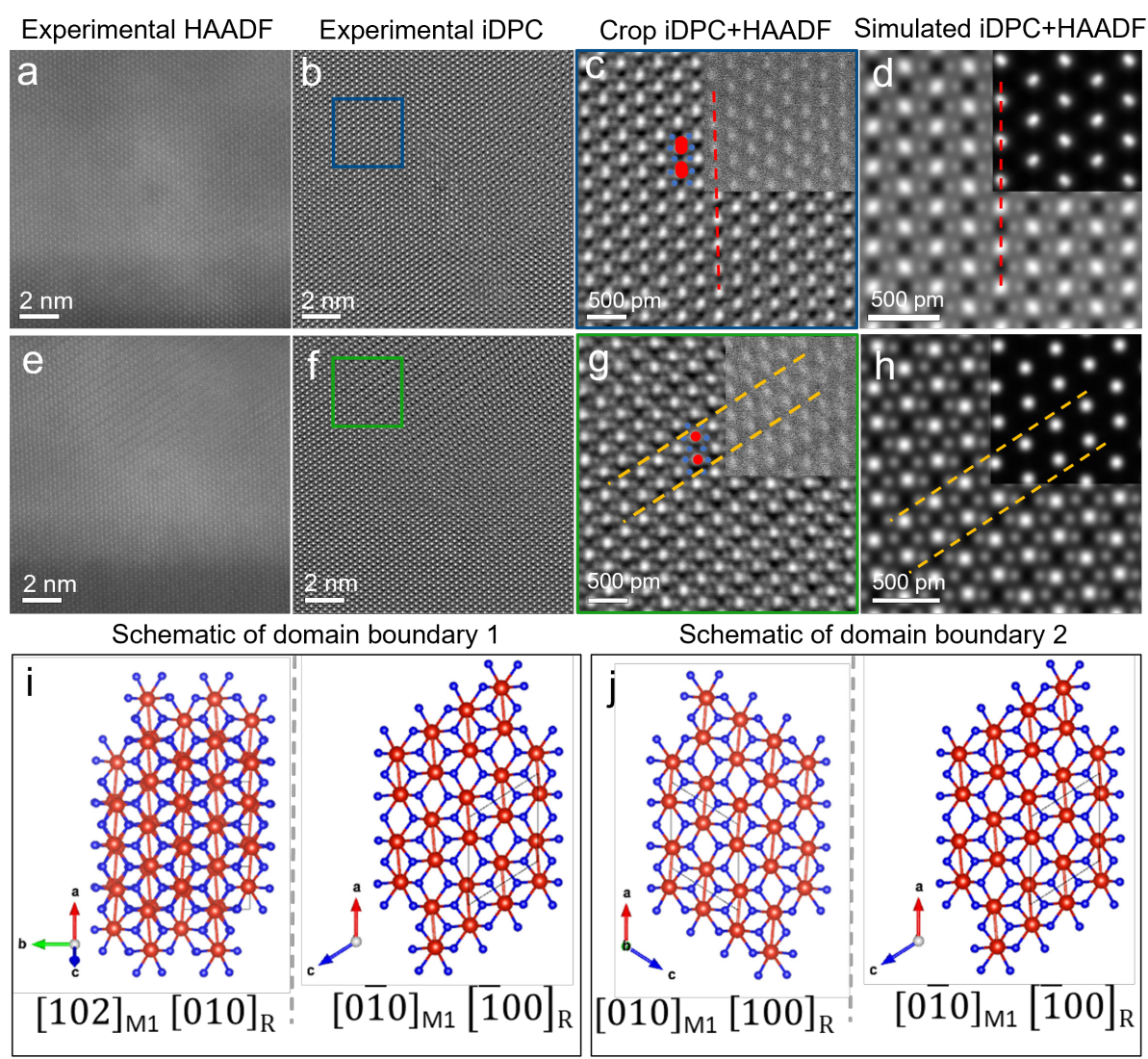

**Figure 9**. HAADF and iDPC of domains 1 and 2 from Figure 8d are shown in (a-b) and (e-f), with the relevant domains cropped and shown in (c,g) with their corresponding simulations in (d,h) respectively. This gives us the schematic of the domains across the cracks from Figure 8d-e shown in (i,j).

We also show the overview of the cross-section made at 45° to the previous lamella in Figure 8c, where we see the cracks at a skew angle as our viewing axis is now [100]/]010] with the c-axis being the vertical direction in the images. Observing the structures around the cracks, we notice different sets of orientations from the inset FFTs of Figure 8d-e. In Figure 9a-b and e-f, we show the HAADF and iDPC images of the two domains from Figure 8d, from which the domains have been cropped in Figure 9c (Crop#1 with the inset showing the HAADF image and the main image being the iDPC image) and 9g (Crop#2 again with the inset showing the HAADF image and the main image being the iDPC image). In the first of these domains, we clearly see V atoms are aligned along c-axis marked with a red dashed line. These V atoms are also slightly elongated along c-axis as compared to the Ti atoms of the substrate. Also, as discussed previously, the V-O bonds perpendicular to the c-axis are equal, and they differ from the other four V-O bonds, which are also equal. This matches excellently to the $[102]_{M1}$ schematic model as shown in Figure 9i and the simulations of the iDPC with inset HAADF are shown in Figure 9d. On the other hand, in the second domain, we do not see the V atoms aligned along c-axis of the substrate as the Ti atoms are. Instead they align along the c-axis of the film marked with yellow dashed lines and therefore, have a structure with dimerized V-V planes. Consequently, all the 6 V-O bonds are now unequal which we find to match with the $[0\bar{1}0]_{M1}$ orientation which is confirmed by simulations shown in Figure 9h. The atomic structure observations for the first domain boundary (Fig. 8d) are summarized in the schematic structure shown in Figure 9i. The results of a similar analysis of the second domain boundary shown in Figure 8e are summarized in the schematic shown in Figure 9j. Here it can be seen that both domains have the dimerized V-V planes, but there dimerization direction is mirrored(twinned) in the domain boundary. So, here we have the $[010]_{M1}$ and the $[0\bar{1}0]_{M1}$ orientations for both domains similar as observed for the thin domain boundary lamella observed in $VO_2$ on the $TiO_2(101)$ substrate in Figure S8(b). Note that, unlike the $[010]$ and $[0\bar{1}0]$ domains, the $[102]$ and $[\bar{1}0\bar{2}]$ domains are identical. So, three types of domains can be distinguished when we view along the $TiO_2[100]$ (or equivalent $TiO_2[010]$).

So while from $[110]_R$ zone-axis, we see the structure completely epitaxial and homogenous across the lamella, from $[100/010]_R$ zone-axis, we always see different orientations of $VO_2$ on the two sides of the cracks. These observations validate previous studies of the cracks developing as a stress relaxation mechanism[44,45] - all these grains have exactly the same rutile structure above transition temperature, but as the thin film cools down, M1 $VO_2$ grains with different orientations nucleate and propagate until they meet resulting in stress which is only able to relax due to the formation of cracks. Overall, the presence of deep cracks in samples on (001) substrate results in poor MIT characteristics, and thus $VO_2$ on this substrate is not suitable for electronic applications unless the stress can be tuned by external mechanisms like doping, or caused to relax with features other than cracks. Such an example does occur for the $VO_2$ film on the $TiO_2(101)$substrate where despite very similar stress conditions due to similarly oriented domains, we observe intricate domain boundary structures instead of cracks. These domain boundary structures involve a thin lamella running inclined through the film at angle corresponding to a high resolved shear stress similar to stress relaxation in a slip system (or with deformation twinning), which enables plastic deformation.[59] (When a boundary would run only perpendicular to the interface and surface, there is no resolved shear stress on the boundary plane and it cannot assist in any plastic deformation and the only response is cracking.) Subsequently, without any cracks the MIT in $VO_2$ on $TiO_2(101)$ remains of high quality, and this substrate should be highly suitable for electronic applications.

## CONCLUSIONS

In summary, we have scrutinized the microstructure of $VO_2$ films grown on different substrates using in particular advanced STEM characterization, also allowing imaging of oxygen atomic columns, and have correlated these microstructures with the observed metal-insulator transition (MIT) properties as measured using van-der-Pauw sheet resistance measurements. Based on our analysis, we conclude that $VO_2$ films grown on c-sapphire consist of very small columnar domains (with column diameters of only 5-10 nm) leading to an MIT spread over a wide temperature range. The monoclinic (M1) $VO_2$ films grown on rutile $TiO_2$ contain micrometer sized domains having boundaries along (110) planes. Cracks between domains are observed on the (001) substrate, but not on the (101) substrate. It is expected that the $VO_2$ films grown with the rutile structure at elevated temperature are fully lattice matched with the rutile $TiO_2$ and that the cracks develop during cooling to room temperature when the rutile to M1 phase transition occurs in the $VO_2$ films. An intricate domain boundary structure is observed in the $VO_2$ on $TiO_2(101)$ that seems to be responsible for the plastic deformation preventing cracking. As a result, we get limited electrical conductivity across the film and very poor MIT characteristics for $VO_2$ on $TiO_2(001)$, but excellent MIT characteristics for $VO_2$ on $TiO_2(101)$, which is therefore a promising platform for electrical devices.

At 500 °C hardly any $VO_2$ film grows on $TiO_2$, but instead a sizeable interdiffusion layer (~15 nm thick) is formed, i.e. where V diffuses into rutile $TiO_2$. Below 400 °C, the intermediate layer is ~3-4 nm thick and further reduction in growth temperature leads to an increase in $VO_2$ film thickness, but with no significant reduction in interdiffusion layer thickness. These results show that the MIT of the $VO_2$ top film contains an additional tunable parameter by varying the thickness and composition of the interdiffusion layer. As a next step our goal is to perform comprehensive STEM-iDPC(+HAADF) analysis to monitor the atomic structure changes of the $VO_2$ by in-situ heating and cooling during the MIT.

## EXPERIMENTAL SECTION

### Pulsed Laser Deposition

Epitaxial $VO_2$ films were grown on single-crystal c-cut sapphire (0001), rutile $TiO_2$ (101) and (001) substrates using pulsed laser deposition (PLD). A KrF excimer laser ($\lambda$ = 248 nm) was focused on a Vanadium metal target (99.9% pure) with a fluence of ~2.5 $Jcm^{-2}$ and pulse rate of 10 Hz, in a vacuum chamber kept at a base pressure lower than $10^{-7}$ mbar throughout the deposition. Growth parameters like the number of pulses, substrate temperature and oxygen pressure ($P_{O2}$) were varied, resulting in stoichiometric $VO_2$ films of varying thickness and quality.

**Van der Pauw sheet resistance measurements**

The Van der Pauw technique was used to measure the sheet resistance of the thin films. This method can be applied for any thin film of arbitrary shape which is of uniform thickness and has no holes, and the resistance so obtained is not affected by contact resistances.[60] Four probes of silver were placed at the perimeter (as close to the boundary as possible) and electrical measurements were done using a Keithley setup to obtain the horizontal ($R_{horizontal}$) and vertical resistances ($R_{vertical}$), from which the sheet resistance ($R_s$) was calculated.

**X-ray Diffraction**

2θ-ω scans were performed using a Panalytical X'pert Pro MRD X-ray diffractometer using a normal stage at room temperature over a wide range of 2θ from 20° to 100°, and then with a heater stage at 2θ values around major substrate peaks to determine the crystal structure of the film before transition (at room temperature), and after transition (at 150 °C). Reciprocal space maps (2θ-ω two-axis measurements) around the major peaks of the substrates were also recorded to find out the epitaxial relationship of the film in-plane on the substrate.

**Specimen preparation for microscopy**

A Helios G4 CX Dual-Beam system with a Ga focused ion beam (FIB) was used to prepare cross-section lamella from the samples. A clean area (15μm x 1.5μm) was carefully chosen on the thin film surface from which the cross-section will be prepared. A protective coating must be deposited over the selected area to prevent ion radiation damage and sample damage or amorphization. Since ion-beam platinum deposition can also damage the very top surface of the sample (~50 nm, which already exceeds our sample thickness), electron-beam Carbon and electron-beam Platinum was deposited first at a very slow rate. Bulk ion milling was performed to reveal the shape of the lamella, and intermediate milling was done with a small tilt of ± 1.5° to thin the lamella to a thickness of around 1-1.5 μm. A U-shaped cut was applied to make the lamella almost free except at one side, and then the lift-out procedure was performed using a manipulator. The lamella was welded to a TEM grid and was ready for subsequent thinning. Thinning was performed on a Peltier-cooled chip, initially with a high voltage of 30 kV up to a thickness of ~100 nm. Low kV (5 kV and 2 kV) thinning was then done to remove the surface amorphization until lamella became electron transparent.

**Scanning transmission electron microscopy (STEM)**

The lamellas prepared were cleaned using plasma for 5-10 minutes before inserting in the microscope column of a probe- and image-corrected ThermoFisher Scientific ThemisZ S/TEM equipment operated at 300 kV, for STEM imaging and elemental mapping. The STEM alignments were done on the amorphous Platinum top layer. Aberrations were corrected up to the fourth order using the TEAM software, and fine-tuning of two-fold condenser astigmatism was done using Sherpa software in order to achieve a resolution of around 64 pm. Defocus values were generally set to negative values from 0 to -10 nm for imaging. For iDPC imaging,[48,61] the gain of the four segments was adjusted to equal intensity background levels to get no deflection of the beam through the amorphous Pt layer.

A semi-convergence angle of ~21-25 mrad was used to get a small depth of focus. The pixel dwell-time was varied between 1 and 10 μs, with the image resolution as 1024x 1024, 2048x 2048, or 4096x4096 pixels. The images obtained were filtered using either Radial Wiener filter or Gaussian filter to get HAADF and iDPC images used for analysis.

**Multi-slice Simulations**

Models of monoclinic and rutile phase of $VO_2$ with the parameters given in supporting information Table 4 and 5, were made using VESTA and exported as cif files. These models were then used in the Dr. Probe software[62] for STEM HAADF and iDPC image simulations of the structures along different orientations. For the simulations, all aberrations except for defocus (kept at -4 nm) were neglected, and the other microscope parameters were set to experimental values (see supporting information Table 7).

## ASSOCIATED CONTENT

Supporting Information is available from the ACS Nano website.

- Supporting Information (docx)

The data that support the findings of this study are available from the corresponding author upon reasonable request.

The authors declare no conflict of interest.

## AUTHOR INFORMATION

### Corresponding Authors
*atul061094@gmail.com
*b.j.kooi@rug.nl

## Author Contributions

A.A. and B.J.K. conceived the idea. A.A. fabricated the thin films and performed all characterization measurements. Training for pulsed laser deposition and optimization of thin films was done with the help of H.Z. STEM measurements were done by P.K. on c-sapphire and by A.A. on the other substrates along with M.A. who was also critical for the analysis of STEM data. A.A. performed the simulations and wrote the paper, which was primarily edited by B.J.K. with critical inputs from the other authors.

## ACKNOWLEDGEMENTS

The authors would like to acknowledge the support of Prof. Dr. Beatriz Noheda from the Zernike Institute for Advanced Materials, University of Groningen, and her group members Jacob Baas and Dr. Qikai Guo.

# TABLE OF CONTENTS

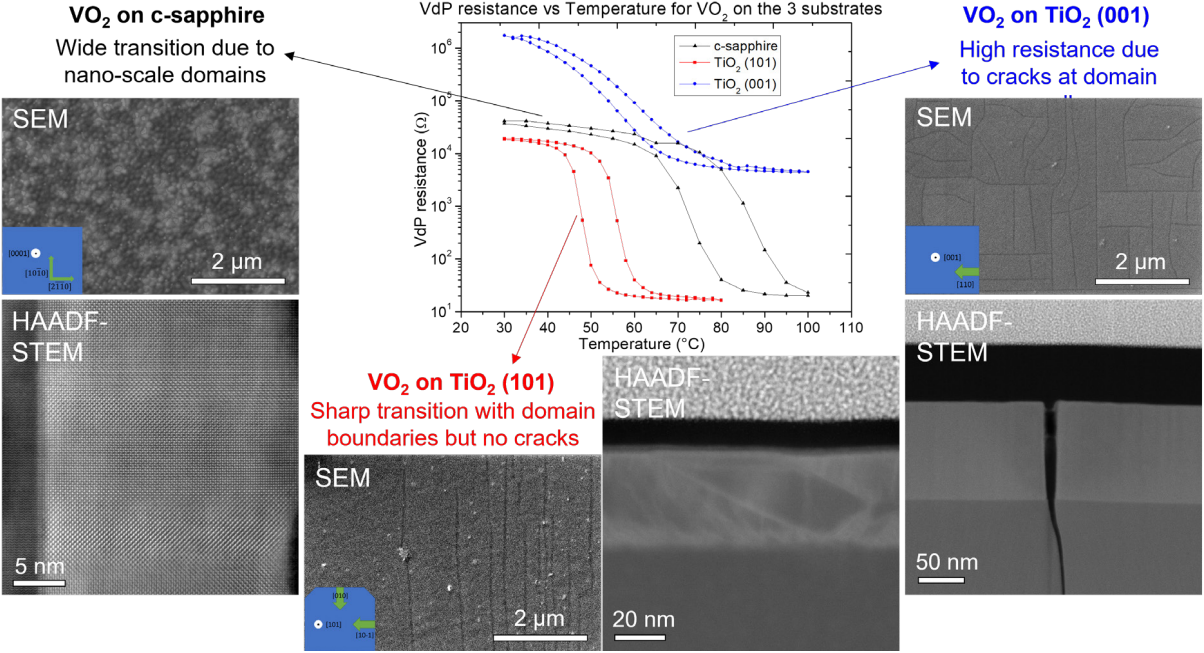